\documentclass[aps,pra,amssymb,amsfonts,floatfix,showkeys]{revtex4}
\usepackage{graphicx}

\begin{document}

\title{Dynamical Manifestations of Quantum Chaos: Correlation Hole and Bulge}

\author{E. J. Torres-Herrera$^{1}$ and Lea F. Santos$^{2}$}

\affiliation{$^{1}$Instituto de F{\'i}sica, Benem\'erita Universidad Aut\'onoma de Puebla, Apt. Postal J-48, Puebla, Puebla, 72570, Mexico \\
$^{2}$Department of Physics, Yeshiva University, New York, New York 10016, USA}


\begin{abstract}
A main feature of a chaotic quantum system is a rigid spectrum where the levels do not cross. We discuss how the presence of level repulsion in lattice many-body quantum systems can be detected from the analysis of their time evolution instead of their energy spectra.  This approach is advantageous to experiments that deal with dynamics, but have limited or no direct access to spectroscopy. Dynamical manifestations of avoided crossings occur at long times. They correspond to a drop, referred to as correlation hole,  below the asymptotic value of the survival probability and by a bulge above the saturation point of the von Neumann entanglement entropy and the Shannon information entropy. In contrast,  the evolution of these quantities at shorter times  reflect the level of delocalization of the initial state, but not necessarily a rigid spectrum.  The correlation hole is a general indicator of the integrable-chaos transition in disordered and clean models and as such can be used to detect the transition to the many-body localized phase in disordered interacting systems. 
\end{abstract}
%%%%%%%%%%%%%%%%%%%%%%%%%%%

\maketitle

\section{Introduction}

Quantum chaos refers to specific properties of the spectrum and the eigenstates of a system, most notably to correlations between the eigenvalues that result in avoided crossings~\cite{MehtaBook,Brody1981,Guhr1998,ZelevinskyRep1996,StockmannBook}. In the context of nuclear physics, the analysis of the statistical properties of nuclear spectra revealing level repulsion dates back to Wigner's works~\cite{Wigner1951P,Wigner1958} and was soon extended to other complex systems, such atoms, molecules and condensed matter models. From a dynamical perspective, the onset of quantum chaos has been associated with very fast relaxation processes. Nevertheless, fast time evolution is found also in non-chaotic systems. This paper is devoted to the description of dynamical quantities that can unambiguously identify the presence of level repulsion in isolated finite lattice many-body quantum systems. 

Strongly perturbed isolated finite many-body quantum systems equilibrate in a probabilistic sense. After a transient time, the dynamics saturates and the observables fluctuate around their infinite-time average, remaining very close to this average value for most of the time.
In an attempt to distinguish quantum motion in chaotic and regular systems, Peres~\cite{Peres1984} argued that after the saturation of the evolution of the Loschmidt echo, the fluctuations in chaotic models should be smaller than in regular systems. More recently, however, it has been shown that after equilibration the amplitudes of the fluctuations of various observables  decrease with system size in a very similar way for chaotic and also integrable systems with interactions~\cite{Reimann2008,Short2011,Zangara2013,KiendlPRL2017}. In terms of the values reached by the observables after equilibration, comparisons between infinite-time averages and thermodynamic averages have been used to determine the onset of thermalization and by extension of quantum chaos~\cite{Borgonovi2016,Alessio2016}, although the results are highly dependent on the initial state~\cite{Rigol2012,Torres2013}. In contrast to these approaches, we focus here on evident manifestations of avoided crossings that emerge before the equilibration of the system.

Quantum chaos has been linked with the linear growth in time of the von Neumann entanglement entropy up to saturation~\cite{Eisert2010}. This behavior is analogous to what happens to the Shannon information entropy~\cite{Torres2016Entropy,TorresAnnPhys}, which has the advantage of not requiring any partial trace of the system. However, the linear growth of these entropies is observed also in interacting integrable models~\cite{Balachandran2010,Santos2012PRL,Santos2012PRE,Torres2016Entropy,TorresAnnPhys}.  Similarly, quantum chaos has been connected with the initial exponential decay of the survival probability and Loschmidt echo~\cite{Jacquod2001,Cucchietti2002,Prosen2002,Weinstein2003}, but the same is verified in integrable models~\cite{TorresProceed} and decays even faster are found in both regimes~\cite{Flambaum2001a,KotaBook,Torres2014PRA,Torres2014NJP,Torres2014PRAb,TorresKollmar2016}.

In one-dimensional interacting systems with onsite disorder, spatial localization due to large disorder occurs in parallel with the disappearance of level repulsion. Dynamical features at the vicinity of the delocalized-localized transition, which encompass the logarithmic growth of both entanglement and Shannon entropies~\cite{Znidaric2008,Bardarson2012,LuitzPRL2017,TorresAnnPhys} and the power-law decay of the survival probability with exponents smaller than 1 \cite{Torres2015,TorresAnnPhys,Torres2016BJP}, may then be taken as signatures of the chaos-integrable transition. But this is a very particular example, where both transitions coincide. We seek for means to differentiate chaos from integrability in general scenarios, including also clean models and disordered systems in the proximity of spatially delocalized integrable points.

The power-law decay of the survival probability at long times is not exclusive to systems in the vicinity of the transition to the many-body localized phase. This behavior is inevitable for any quantum system, but its source as well as the values of the power-law exponents vary~\cite{Tavora2016,Tavora2017}.  One ever-present cause of the power-law decay is the unavoidable bounds in the energy spectrum of quantum systems~\cite{MugaBook,Campo2016}. In lattice many-body quantum systems with two-body interactions, the spectrum energy bound can lead to power-law exponents close to 2 \cite{Tavora2016,Tavora2017}. Such large exponents indicate chaotic initial states~\cite{Tavora2016,Tavora2017}, but they are still not explicit signatures of level repulsion.

Unambiguous dynamical manifestations of level repulsion occur at yet longer times, after the power-law decay and before equilibration. It shows up in the form of a drop in the value of the survival probability below its saturation point, a phenomenon known as correlation hole~\cite{Leviandier1986}. As we discuss here, it also leads to a minor bulge above the saturation values of the Shannon and entanglement entropies. The correlation hole reflects the correlations of the eigenvalues of complex systems. It is the dip of the spectral form factor discussed in~\cite{Cotler2017}. It has also been investigated in molecules~\cite{Leviandier1986,Pique1987,Guhr1990,Lombardi1993,Michaille1999,Alhassid2006}, random matrices~\cite{Hartmann1991,Delon1991,Alhassid1992,Gorin2004,Leyvraz2013}, microwave billiards~\cite{Kudrolli1994,Alt1997} and disordered spin models~\cite{TorresAnnPhys}. Here, we extend these studies to clean and disordered spin-1/2 models. 

We stress that in this work we equate the term quantum chaos with level repulsion, more precisely with the Wigner-Dyson distribution of the spacings between neighboring levels~\cite{MehtaBook,Brody1981,Guhr1998}. An alternative approach is to view quantum chaos as the emergence of chaotic eigenstates. The latter refers to states with a very large number of uncorrelated components, which may occur even in systems that do not show level repulsion~\cite{Borgonovi2016}. This leads to generic dynamical behaviors that exhibit very fast relaxation processes. This second viewpoint is inspired by early works from Chirikov~\cite{Chirikov1985,Chirikov1997}.

This paper is organized as follows. Section~2 presents the models and dynamical quantities studied. Section~3 compares the dynamics of integrable and chaotic models without random disorder, using for that the evolution of the entanglement entropy,  the Shannon entropy, and the survival probability. Section~4 shows how the correlation hole can be used to indicate the transition to a many-body localized phase. Section 5 presents the Conclusions.

\section{Spin-1/2 Models and Quantities Analyzed}

We investigate a one-dimensional spin-1/2 system with an even number $L$ of sites. The Hamiltonian is given by
\begin{eqnarray}
H &=& \varepsilon_1 J S_{1}^z  + \varepsilon_L J S_{L}^z + d J S_{L/2}^z + \sum_{k=1}^{L} h_k J S_k^z \nonumber \\
&+& J\sum_{k} \left( S_k^x S_{k+1}^x + S_k^y S_{k+1}^y +\Delta S_k^z S_{k+1}^z \right) \nonumber \\
&+& \lambda J\sum_{k} \left( S_k^x S_{k+2}^x + S_k^y S_{k+2}^y +\Delta S_k^z S_{k+2}^z \right).
\label{hamXXZ}
\end{eqnarray}
Above, $\hbar =1$ and $S^{x,y,z}_k =\sigma^{x,y,z}_k/2$ are spin operators acting on site $k$, $\sigma^{x,y,z}_k$ being Pauli matrices. Three defects are created by applying three different local static magnetic fields in the $z$-direction on the first, last, and middle sites. These fields lead, respectively, to the following Zeeman splittings: $\varepsilon_1 J, \varepsilon_L J$, and $d J$. The Zeeman splittings $h_k J$ correspond to onsite disorder caused by additional static magnetic fields; $h_k$ are random numbers from a uniform distribution in $[-h,h]$.  Equation~(\ref{hamXXZ}) contains couplings between nearest neighbor (NN) and next-nearest neighbor (NNN) sites. They include the flip-flop terms and Ising interactions. $J$ is the exchange coupling, $\Delta$ is the anisotropy parameter, and $\lambda$ is the ratio between the NN and NNN couplings. We set $J=1$. The sum for NN (NNN) couplings runs up to $L-1$ ($L-2$) for open boundary conditions and up to $L$ for closed (periodic) boundary conditions. 

Hamiltonian (\ref{hamXXZ}) conserves the total spin in the $z$-direction, ${\cal S}^z = \sum_{k=1}^L S_k^z$, so $[H,{\cal S}^z]=0$. We deal with the sector that has $L/2$ up-spins and dimension ${\cal D}=L!/[(L/2)!]^2$.
 
The purpose of the defects on site $1$ and site $L$ is to reduce finite size effects and break symmetries, such as parity, conservation of total spin, and spin reversal. We consider $\varepsilon_1, \varepsilon_L$ as small random numbers from a uniform distribution in the interval $[-0.1,0.1]$.

\subsection{Spin Models With and Without Random Disorder}
In Sec.~\ref{SecClean}, the results for three models without random onsite disorder ($h=0$), but with $\varepsilon_1, \varepsilon_L \neq 0$, and open boundary conditions are compared. The models are:

\vskip 0.2 cm
(i) For $\Delta \neq 0$ and $d, \lambda=0$, Eq.~(\ref{hamXXZ}) corresponds to the integrable {\bf XXZ model}. Notice that  the integrability is not broken by the addition of border defects~\cite{Alcaraz1987}. We choose $\Delta=0.48$. The level spacing distribution in this case is Poisson, as typical of integrable models, where the eigenvalues are uncorrelated and crossings are not prohibited.

\vskip 0.2 cm
(ii) The system becomes chaotic and shows a Wigner-Dyson distribution when $0<d\lesssim 1$ \cite{Santos2004,Torres2014PRE,Gubin2012}. We fix $d=0.9$ and $\lambda=0$ and refer to it as the {\bf defect model}.

\vskip 0.2 cm
(iii) Chaos also emerges when $0<\lambda \lesssim 1$ \cite{Hsu1993,Kudo2005,Santos2009JMP,Gubin2012}. We choose $\lambda=1$ and $d=0$ and denote this case as the {\bf NNN model}.

\vskip 0.2 cm
In Sec.~\ref{SecLoc}, we study the {\em disordered model} of spins 1/2 with $h\neq 0$, $\Delta=1$, $\varepsilon_1, \varepsilon_L,d, \lambda =0$,   and periodic boundary conditions. As $h$ increases from zero to $h\lesssim 1$, the level spacing distribution changes from Poisson to Wigner-Dyson~\cite{Avishai2002,Santos2004} and the eigenstates become even more delocalized in space~\cite{SantosEscobar2004}. As the disorder strength is further increased, the distribution transitions from Wigner-Dyson back to Poisson and the eigenstates become more localized in space~\cite{Santos2004,SantosEscobar2004,Dukesz2009}.

\subsection{Basis and Initial States}
When studying localization in space, it is natural to choose the site-basis vectors, as we do in this work. They are also known in quantum information theory as computational-basis vectors. These states have on each site a spin that either points up or down in the $z$-direction, as for example $|\downarrow  \uparrow \downarrow \uparrow \downarrow \uparrow \downarrow \uparrow \ldots \rangle_z$. We notice, however, that further insights on the interplay between interaction and disorder may be gained by analyzing the eigenstates also in other basis vectors, such as those corresponding to the eigenstates of the XXZ model~\cite{Dukesz2009}.

The site-basis vectors are denoted by $|\phi_n \rangle$. They are the initial states, $|\Psi(0) \rangle = |\phi_{n=ini} \rangle $, that we use in the analysis of the system dynamics.

\subsection{Dynamical Quantities}
We study the evolution of the von Neumann entanglement, Shannon information entropy, and survival probability.

\subsubsection{Entanglement Entropy}

The von Neumann entanglement entropy, $S_{vN}$, is obtained by separating the system in subsystems A and B and then performing the partial trace of one of the two~\cite{Amico2008}. The entanglement entropy is the von Neumann entropy of the reduced density matrix $\rho_A = \text{Tr}_B [\rho ]$, where $\rho$ is the density matrix of the total system. We divide the chain in two equal sizes, so the dimension of $\rho_A$ is ${\cal D}_{A}=2^{L/2}$.

The system is initially in the product state $\rho(0) = |\Psi(0) \rangle \langle \Psi(0) | =  |\phi_{ini} \rangle \langle \phi_{ini}|$, so $S_{vN} (0)=0$. As time passes, the amount of entanglement grows as quantified by
\begin{equation}
S_{vN} (t)= - \text{Tr} \left[ \rho_A (t) \ln \rho_A (t) \right] .
\label{SvN}
\end{equation} 

\subsubsection{Shannon Entropy}

The Shannon information entropy, $S_{Sh}$, is often used to measure the level of delocalization of the eigenstates in a chosen basis~\cite{ZelevinskyRep1996}. It can also be used to quantify the spreading in time of the initial state on a selected basis. For the site-basis vectors, it is written as
\begin{equation}
S_{Sh}(t)= - \sum_{n=1}^{ {\cal D} } P_n (t) \ln P_n (t)  ,
\label{Sh}
\end{equation}
where $P_n (t) =  \left| \langle \phi _n | e^{ - iHt} |  \phi _{ini} \rangle  \right|^2$ and ${\cal D}$ is the dimension of the subspace considered. 

We note that the Shannon entropy when written in the energy eigenbasis is now commonly referred to as diagonal entropy~\cite{Polkovnikov2011,Santos2011PRL}. Comparisons between the diagonal entropy and entanglement entropy in the context of thermalization can be found in~\cite{Santos2012PRER}.

\subsubsection{Survival Probability and Correlation Hole}
\label{Sec:SP}

The survival probability, $P_{ini}=\left| \langle \Psi(0) | \Psi(t) \rangle \right|^2$, is the probability for finding the system in its initial state later in time. For $|\Psi(0)\rangle = |\phi_{ini}\rangle$, it is given by 
\begin{equation}
P_{ini} (t) = \left| \langle \phi_{ini} | e^{-i H t} | \phi_{ini} \rangle \right|^2 = \left|\sum_{\alpha}  \left| C_{ini}^{(\alpha)} \right|^2 e^{-i E_{\alpha} t}  \right|^2 = \left| \int \rho_{ini} (E) e^{-i E t} dE  \right|^2, 
\label{P0}
\end{equation}
where $C^{(\alpha)}_{ini} = \langle \psi_{\alpha} | \phi_{ini} \rangle $ is the overlap between the initial state and the eigenstates $|\psi_{\alpha} \rangle$ of the Hamiltonian $H$ that evolves $|\Psi(0)\rangle$, $E_{\alpha}$ are the eigenvalues of $H$, and $\rho_{ini} (E) =\sum_{\alpha}  | C_{ini}^{(\alpha)}|^2 \delta (E - E_{\alpha }) $ is the energy distribution weighted by the squared overlaps $|C^{(\alpha)}_{ini}|^2$. This distribution is known as the local density of states (LDOS) or strength function. 

The survival probability is the absolute square of the Fourier transform of the LDOS. If one has detailed information about the LDOS, one should be able to predict the evolution of $P_{ini} (t)$. Equivalently, $P_{ini} (t) $ is the Fourier transform of the spectral autocorrelation function, $G(E)$, that is
\begin{eqnarray}
P_{ini} (t) &=& \int G(E) e^{-i E t} dE , 
\label{intP0GE}
\\
G(E) &=& \sum_{\alpha}| C^{(\alpha)}_{ini}|^4 \delta( E - E_{\alpha} ) +  \sum_{\beta \neq \alpha}| C^{(\beta)}_{ini}|^2 |C^{(\alpha)}_{ini}|^2
\delta( E - ( E_{\alpha} - E_{\beta} ) )  .
\label{P0GE}
\end{eqnarray}
It is clear from Eqs.~(\ref{intP0GE}) and~(\ref{P0GE}) that the dephasing of the initial state depends on both the initial state, through the overlaps $|C^{(\alpha)}_{ini} |^2$, and the spacings between all energy levels.

The first term of $G(E)$ leads to the infinite-time average of the survival probability $\overline{P_{ini}} = \sum_{\alpha} | C^{(\alpha)}_{ini}|^4$. It is larger than zero in finite systems. The lowest values are reached by full random matrices (FRM), which are matrices filled with random numbers whose sole constraint is to satisfy the symmetries of the system they represent~\cite{Guhr1998}. For FRM of Gaussian Orthogonal Ensembles (GOE), $\overline{P_{ini}} \simeq 3/{\cal D}$ \cite{ZelevinskyRep1996,Torres2016Entropy}, where ${\cal D}$ is the dimension of the random matrix.

The second term of $G(E)$ determines the decay of $P_{ini} (t) $ and the fluctuations around the infinite-time average. In the case of FRM, it factorizes into a term depending only on the overlaps and one depending only on the eigenvalues, as
\begin{equation}
\sum_{\beta \neq \alpha} \left\langle  |C^{(\beta)}_{ini} |^2 |C^{(\alpha)}_{ini} |^2  \right\rangle_{FRM}
\left\langle \delta( E - ( E_{\alpha} - E_{\beta} ) )  \right\rangle_{FRM} .
\end{equation}
Above, $\langle . \rangle_{FRM}$ denotes the average over an ensemble of FRM.

The average over the distribution of level spacings leads to
\begin{equation}
\left\langle \delta( E - ( E_{\alpha} - E_{\beta} ) )  \right\rangle_{FRM} = \frac{1}{ {\cal D} ({\cal D}-1)} \int 
\delta( E - ( E_1 - E_2 ) )  R_2(E_1 , E_2 ) dE_1 dE_2 ,
\end{equation}
where $R_2(E_1 , E_2 )$ is the Dyson's two-level correlation function~\cite{MehtaBook}. $R_2(E_1 , E_2 )$ gives the probability for finding a level around the energies $E_1$ and $E_2$. This function can be written as $R_2(E_1 , E_2)  = R_1(E_1) R_1(E_2) - T_2(E_1,E_2) $, where $R_1(E)$ is the density of states and $T_2(E_1,E_2) $ is the two-level cluster function~\cite{MehtaBook}.

The Fourier transform of $T_2(E_1,E_2)$ is non-zero only in systems that show level repulsion, being zero if the level spacing distribution is Poisson~\cite{Guhr1998}. The Fourier transform of $T_2(E_1,E_2)$ is directly related to the level number variance~\cite{Guhr1998,StockmannBook}, a quantity that measures the level of rigidity of the spectrum. Contrary to signatures of chaos associated with the spacings of neighboring levels, the level number variance detects long-range correlations between the eigenvalues. Correlations between energy levels are the source of the drop of $P_{ini} (t) $ below $\overline{P_{ini}}$, which is known as correlation hole~\cite{Leviandier1986,Pique1987,Guhr1990,Lombardi1993,Michaille1999,Alhassid2006,Hartmann1991,Delon1991,Alhassid1992,Gorin2004,Leyvraz2013,Kudrolli1994,Alt1997}. The correlation hole occurs at long times and disappears as we approach the Heisenberg time (the inverse of the mean level spacing~\cite{Garcia2016}). In the particular case of FRM from GOE, the minimal value reached by $P_{ini}(t)$ due to the correlation hole is $2/{\cal D}$ \cite{Alhassid1992}.

In FRM, $R_1(E)$ and $\rho_{ini} (E)$ coincide, both having a semicircular form~\cite{Torres2014PRA,Torres2014NJP,Torres2014PRAb}. The similarity between density of states and LDOS holds also in realistic many-body models with two-body interactions when the initial state is very delocalized in the energy eigenbasis. However, in this case, the shape of both distributions is Gaussian~\cite{Torres2014PRA,Torres2014NJP,Torres2014PRAb}. The shape of the LDOS determines the behavior of the survival probability at short times. The Fourier transform of a Gaussian envelope results in the Gaussian decay $P_{ini} = \exp ( - \omega_{ini}^2 t^2)$, where
\begin{equation}
\omega_{ini}^2 = \langle \Psi(0) |H^2 |\Psi(0) \rangle -  \langle \Psi(0) |H |\Psi(0) \rangle^2 = \sum_{n \neq ini} \left| \langle  \phi_n|H|\phi_{ini} \rangle \right|^2
\end{equation}
is the square of the width of the LDOS. Notice that this is not simply the quadratic decay that develops at very short-times, $t \ll \omega^{-1}_{ini}$, but a true Gaussian behavior that can hold until the inevitable power-law decay develops~\cite{Tavora2016,Tavora2017}.

\section{Dynamical Manifestation of Level Repulsion}
\label{SecClean}

We compare results for the integrable XXZ model with those for the chaotic defect and NNN models. To reduce finite size effects, we average the results over different realizations of random numbers representing the small border defects $\varepsilon_1$ and $\varepsilon_L$. We also perform averages over initial states chosen at random among all site-basis vectors. Since the density of states is Gaussian, the majority of these states have energy close to the middle of the spectrum. The results after performing all averages are denoted by $\langle . \rangle$. 

\subsection{Growth in Time of the Entanglement and Shannon Entropies}

\begin{figure}[t!]
\centering\includegraphics[width=4.5in]{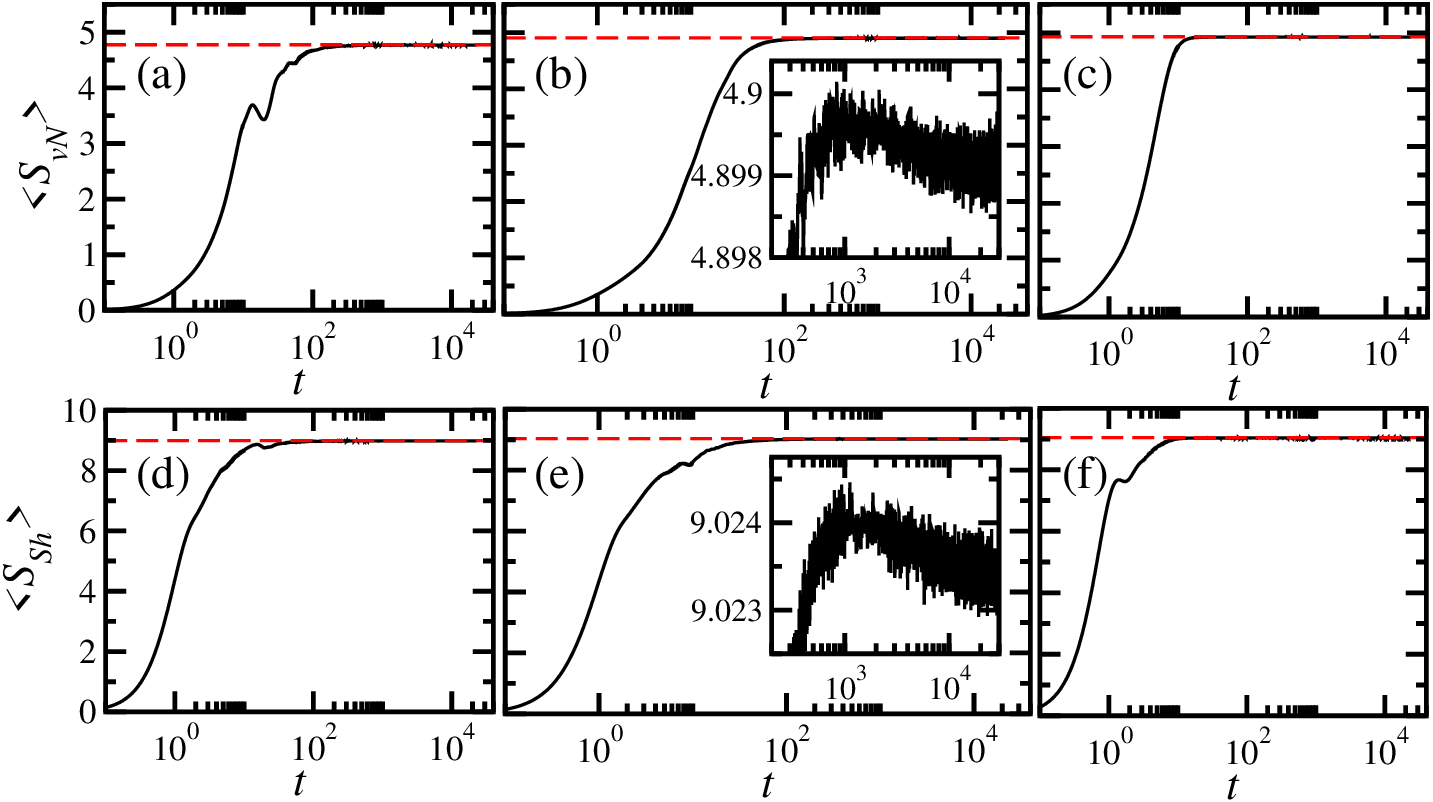}
\caption{Evolution of the entanglement entropy (top panels) and Shannon entropy (bottom panels) for the XXZ (a, d), defect (b, e), and NNN (c, f) models. Dashed horizontal lines give the saturation values. The insets in (b) and (e) are zoom ins of the plots at long times. The parameters are $\Delta=0.48$, $d=0.9$, $\lambda=1$, $h=0$, $L=16$, ${\cal S}^z=0$, ${\cal D}=12\;870$, ${\cal D}_A=256$; open boundaries. Average over 50 initial states corresponding to randomly selected site-basis vectors and average over 20 realizations of random border defects. }
\label{FigEntropy}
\end{figure}

The main panels in Fig.~\ref{FigEntropy} reinforce ideas presented in previous works. They show the evolution of the entanglement entropy (top panels) [Eq.~(\ref{SvN})] and of the Shannon entropy (bottom panels) [Eq.~(\ref{Sh})]  for the XXZ (a,d), defect (b,e) and NNN (c,f) models. As described below, the behaviors are very similar for both entropies and for the integrable and chaotic models. 

For $t \ll \omega^{-1}_{ini}$, the entropies grow quadratically~\cite{Torres2014NJP}, as expected by simply expanding the expressions in Eq.~(\ref{SvN}) and Eq.~(\ref{Sh}). Subsequently, the increase is nearly linear until saturation. The linear increase of the entropies reflects the large number of decay channels available for the initial state, not necessarily the presence of level repulsion. This was stressed in Refs.~\cite{Torres2016Entropy,Santos2012PRL,Santos2012PRE}, where clean models were considered, and also in Ref.~\cite{TorresAnnPhys} for disordered systems. In~\cite{Santos2012PRL,Santos2012PRE}, the Shannon entropy was studied for initial states corresponding to mean-field basis vectors. This allowed for the derivation of analytical expressions following the steps discussed in~\cite{Flambaum2001b}. In~\cite{Torres2016Entropy} both entropies were considered, but only for two specific site-basis vectors, the N\'eel state and the domain wall state. Here, we average the results over 50 different site-basis vectors and also over 20 realizations of random border defects, so the curves are smoother.

%\vspace*{10pt}
%\vspace*{-5pt}

The fact that both entropies lead to very similar behaviors suggests that any of the two can be equivalently used to study nonequilibrium quantum dynamics. In this context, entanglement does not appear to be an essential property~\cite{SantosEscobar2004,Brown2008}. The advantage of the Shannon entropy is to be computationally less expensive, since it does not require the partial trace of the system. It would be interesting, however, to identify which features of the dynamics of many-body quantum systems one entropy can detect that the other cannot.

At first sight, the results for the entropies in the main panels of Fig.~\ref{FigEntropy} seem unable to differentiate integrable from chaotic models. However, the Shannon entropy explicitly contains the survival probability,
\begin{equation}
S_{Sh}(t)= -  P_{ini} (t) \ln P_{ini} (t) - \sum_{n\neq ini} P_n (t) \ln P_n (t)  ,
\label{EqBulge}
\end{equation} 
so one might expect that it could capture some signature of level repulsion. Specifically, in the time interval where the correlation hole takes place for the survival probability, Eq.~(\ref{EqBulge}) suggests that an increase beyond the saturation value for the Shannon entropy could happen. By substantially zooming in the results at long times, we indeed find such bulge. It is visible for both entropies in the chaotic defect model, as shown in the insets of Fig.~\ref{FigEntropy} (b) and (e), but no sign of it appears in the integrable XXZ model. We note that the averages over initial states are necessary to smoothen the curves and reveal the correlation bulge. Averages over more states and for larger system sizes should further reduce the fluctuations.

\subsection{Emergence of the Correlation Hole}

\begin{figure}[t!]
\centering\includegraphics[width=4.5in]{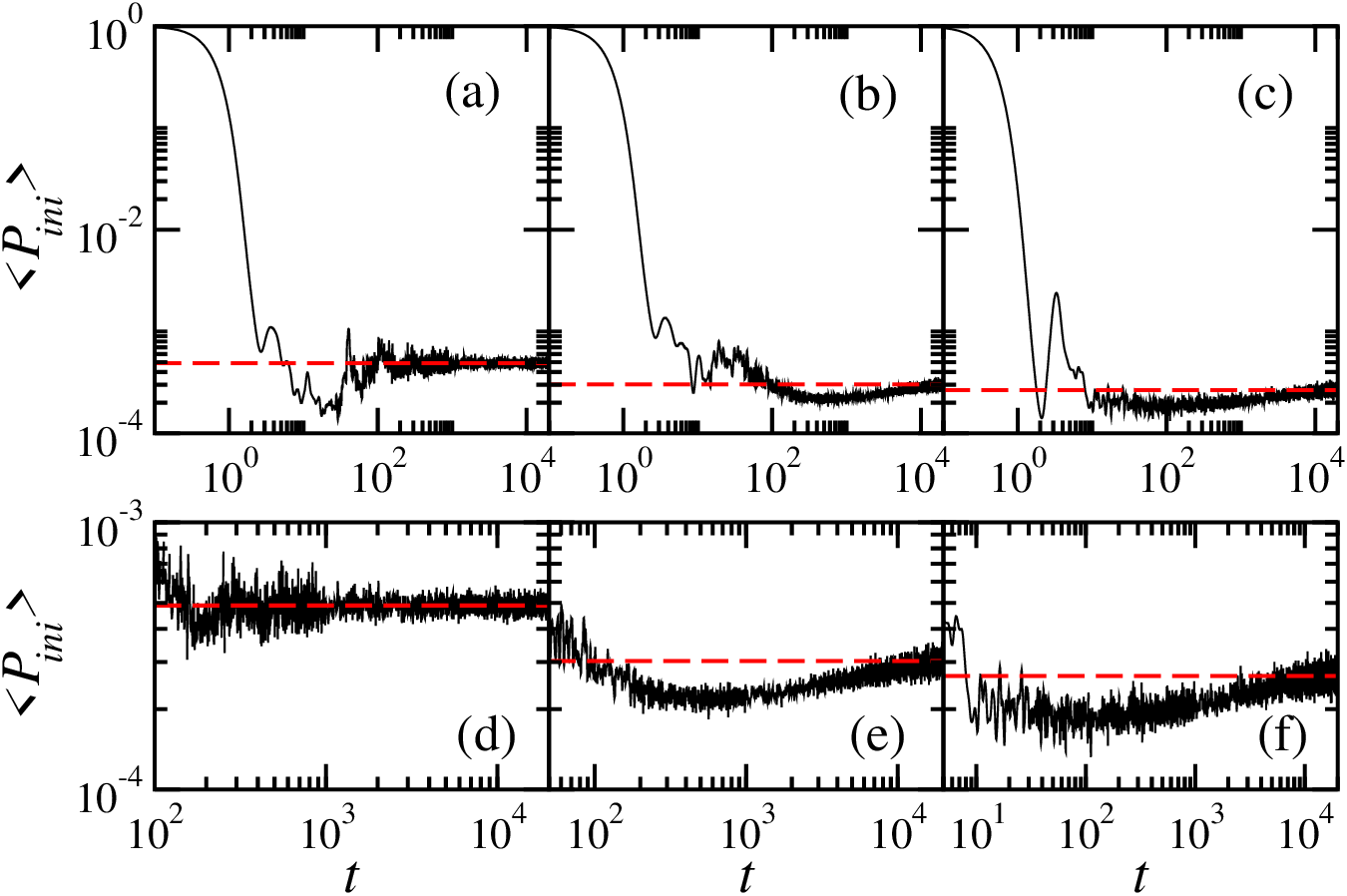}
\caption{Survival probability for the XXZ (a,d), defect (b,e), and NNN (c,f) models. The parameters are $\Delta=0.48$, $d=0.9$, $\lambda=1$, $h=0$, $L=16$, ${\cal S}^z=0$, ${\cal D}=12\;870$. Average over 50 initial states corresponding to randomly selected site-basis vectors and average over 20 realizations of random border defects. }
\label{FigFid}
\end{figure}

The survival probability is a very simple quantity that contains a lot of information about the system and its evolution at different time scales. When systems with two-body interactions are strongly perturbed out of equilibrium, as in the cases where site-basis vectors evolve under the XXZ, defect, and NNN models, the envelope of the LDOS is Gaussian and the initial decay of the survival probability is also Gaussian~\cite{Torres2014PRA,Torres2014NJP,Torres2014PRAb}. This is illustrated in Fig.~\ref{FigFid} for the three models: XXZ (a,d), defect (b,e) and NNN (c,f). This behavior persists up to $t\sim 2$. Between $t\sim 2$ and $t\sim 10$, there are oscillations most likely associated with finite size effects and the energy bounds of the spectrum~\cite{Tavora2016,Tavora2017}. They can lead to values of $\langle P_{ini}(t)\rangle $ below the saturation line, but they are not yet related to the correlation hole. In the examples of Fig.~\ref{FigFid}, these drops are more significant for the clean XXZ and NNN models than for the defect model. 

%\vspace*{-7pt}
%\vspace{-5pt}

Due to the energy bounds in the spectrum, the initial Gaussian decay gives place to a power-law decay at later times. For small system sizes, this behavior is noticeable in disordered models~\cite{Tavora2016,Tavora2017,TorresAnnPhys}, where averages over several disorder realizations and initial states are performed. The power-law decay is not visible in Fig.~\ref{FigFid}, probably because of large finite size effects. However, the small values of the survival probability, $\langle P_{ini}(t) \rangle <\overline{P_{ini}}$, for $t\lesssim 10$ suggest an underneath power-law behavior.  In the vicinity of the point where the decay of the survival probability changes from Gaussian to power-law, there occurs an interference between the two contributions. This causes a phenomenon known as survival collapse~\cite{Fiori2006,Fiori2009,MugaBook} that often results in $P_{ini}(t)<\overline{P_{ini}}$, as indeed confirmed for spin-1/2 systems in Ref.~\cite{Tavora2017}. 

It is for times even longer, $t>10$, that the correlation hole finally develops, first for the NNN model, where the minimum occurs at $t\sim 111$ for the parameters of Fig.~\ref{FigFid},  and later for the defect model, where the minimum is at $t\sim 564$. This difference in time must be caused by the level of rigidity of the spectra of the two models. For the chosen parameters, the spectrum of the NNN model is more rigid than that for the defect model~\cite{Torres2014PRE}.

By comparing Fig.~\ref{FigFid} (e) and the inset of Fig.~\ref{FigEntropy} (e), we notice that the maximum value in the bulge of the Shannon entropy occurs at a time of the same order of magnitude as the time for the minimum of the correlation hole. % at $t\sim 1128$.
After the correlation hole, the survival probability simply fluctuates around the saturation point, $\overline{P_{ini}}$. The fluctuations tend to be smaller in chaotic models [compare Fig.~\ref{FigFid} (d) and Fig.~\ref{FigFid} (e)], but they decrease with system size in a similar way for chaotic and interacting integrable models~\cite{Zangara2013}.

%%%%%%%%%%%%%%%%%%%%%%%%%%%%%%%%%%%%%%%%%%%%%%%%%%
\section{Transition to the Many-Body Localized Phase}
\label{SecLoc}

The characterization of the interacting disordered spin-1/2 model with $h\neq0$ and $\lambda=0$ in terms of level statistics as well as delocalization and entanglement measures was first performed in 2004 \cite{SantosEscobar2004}. As $h$ increases from zero, $0\leq h \leq1$, the system undergoes a transition from integrability to chaos followed in parallel by an increase in the level of delocalization of the eigenstates written in the site-basis. As $h$ further increases above 1, the system undergoes a second transition, now from chaos to integrability along with the spatial localization of the eigenstates. This section focus on the two transition regions, with emphasis on the second one, between the chaotic phase and the spatially localized phase.

In this second transition region, the level statistics is intermediate between Wigner-Dyson and Poisson and the eigenstates are multifractal~\cite{Torres2015}. We deal in this case with extended nonergodic eigenstates that become more correlated as $h$ increases~\cite{TorresAnnPhys}.  These correlations are responsible for the logarithmic growth of the entanglement~\cite{Znidaric2008,Bardarson2012} and Shannon entropies~\cite{TorresAnnPhys}, and for the power-law decays of local observables~\cite{Serbyn2014,Luitz2016},  out-of-time order correlators~\cite{Fan2017}, and survival probability with power-law exponents smaller than 1 \cite{Torres2015,TorresAnnPhys}. It was shown in Refs.~\cite{Torres2015,TorresAnnPhys} that $S_{Sh,vN}(t) \sim D_2 \ln(t)$ and $W_{ini}(t) \propto t^{-D_2}$, where $D_2$ is the fractal dimension obtained from scaling analysis of the participation ratio of the initial state projected into the energy eigenbasis. $D_2$ also agrees with the fractal dimension obtained from scaling analysis of the participation ratio of the eigenstates written in the site-basis vectors.

Since the spatial delocalized-localized transition in disordered models with interactions happens in parallel with the transition from a  Wigner-Dyson  to a Poisson distribution~\cite{SantosEscobar2004}, we can use the disappearance of level repulsion as a signature of the transition to the many-body localized phase. One way to do this is by computing the eigenvalues to directly study level statistics. The other is by analyzing the evolution of the survival probability and how the correlation hole fades away as the disorder strength becomes large. The latter is the approach taken here. As explained in Sec.~2 (c) (iii), the correlation hole exists only in systems that show level repulsion. There is a one to one correspondence between level statistics and the long-time behavior of the survival probability.

%\vspace*{-7pt}
\begin{figure}[t!]
\hspace{-0.7 cm}
\centering\includegraphics[width=5.2in]{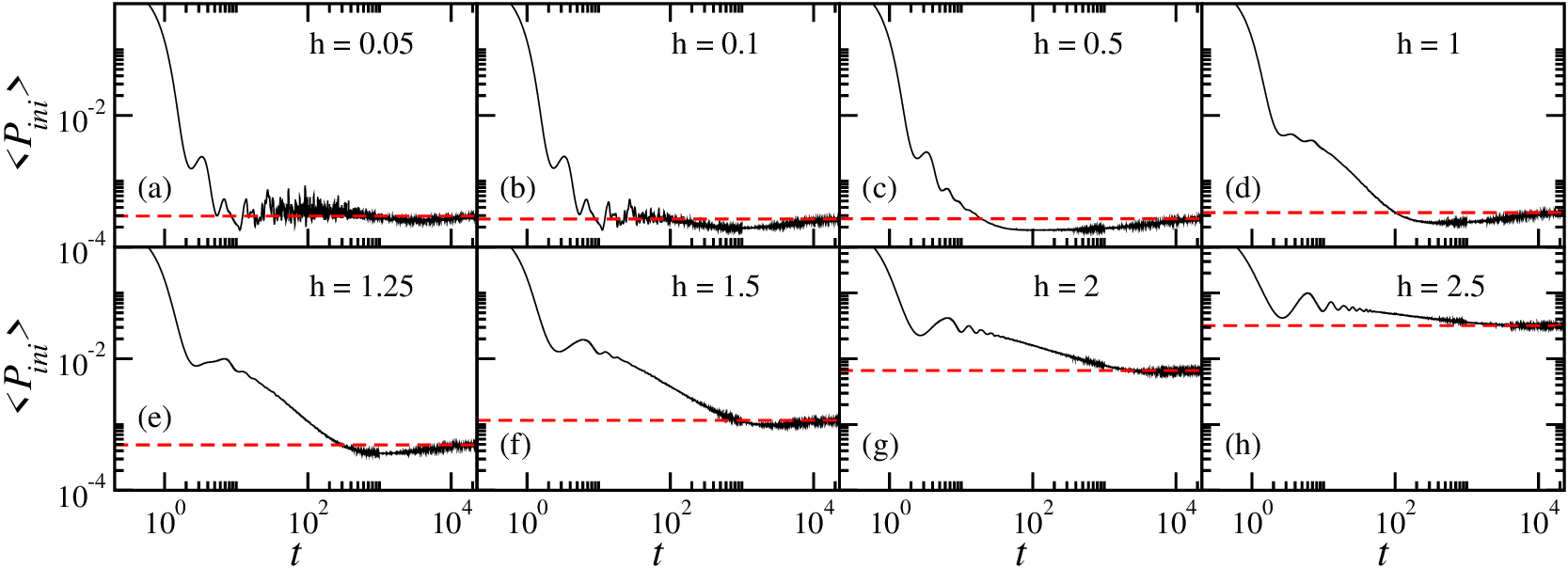}
\caption{Survival probability up to saturation. The values of $h$ are indicated in the figure; $\Delta=1$, $\varepsilon_{1,L}, d, \lambda=0$, $L=16$, ${\cal S}^z=0$. The dashed-lines in (a)-(h) indicate the infinite-time average, $\overline{P_{ini}}$. Average performed over $\sim 10^3$ different initial states corresponding to site-basis vectors with energy in the middle of the spectrum and over $10^2$ disorder realizations.}
\label{FigLoc}
\end{figure}
%\vspace*{-5pt}

Figure~\ref{FigLoc} shows the survival probability for various values of $h$ smaller than the critical point for the transition to spatial localization. The decay is Gaussian at short times and subsequently becomes power-law. For disorder strengths in the intermediate region between the chaotic limit and the many-body localized phase, the Fourier transform of the autocorrelation function in Eq.~(\ref{intP0GE}), $G(E) \propto E^{D_2-1}$, leads to the power-law decay with exponent $D_2<1$. This exponent reflects the level of correlations of the eigenstates~\cite{Torres2015,TorresAnnPhys}. In the chaotic regime, the power-law exponent is larger than 1 and can no longer be explained in terms of the correlations between eigenstates. The power-law decay in this case is associated with the inevitable presence of energy bounds in the spectrum of quantum systems~\cite{Tavora2016,Tavora2017}.  By taking these bounds into account when performing the Fourier transform of a Gaussian LDOS in Eq.~(\ref{P0}), we find the maximum value $2$ for the power-law exponent.

At long times, beyond the power-law decay and before the saturation of $\langle P_{ini}(t) \rangle $, the correlation hole emerges when the spectrum shows level repulsion.  In Fig.~\ref{FigLoc} (a), where the disorder strength is very small and the system is close to the delocalized integrable XXZ model, one hardly sees the hole. It gets deeper as $h$ increases from zero. The maximum level of chaoticity, in the sense of proximity to the GOE Wigner-Dyson distribution, happens at $h\sim 0.5$ for the system size considered here. At this point, the correlation hole is deepest [Fig.~\ref{FigLoc} (c)]. For even larger $h$, the hole starts shrinking once again.  As the level of correlation between the eigenvalues decreases and the spectrum becomes less rigid, the correlation hole gets postponed to later times and fades away, as verified from Fig.~\ref{FigLoc} (c) to Fig.~\ref{FigLoc} (h).

\begin{figure}[t!]
\centering\includegraphics[width=3.7in]{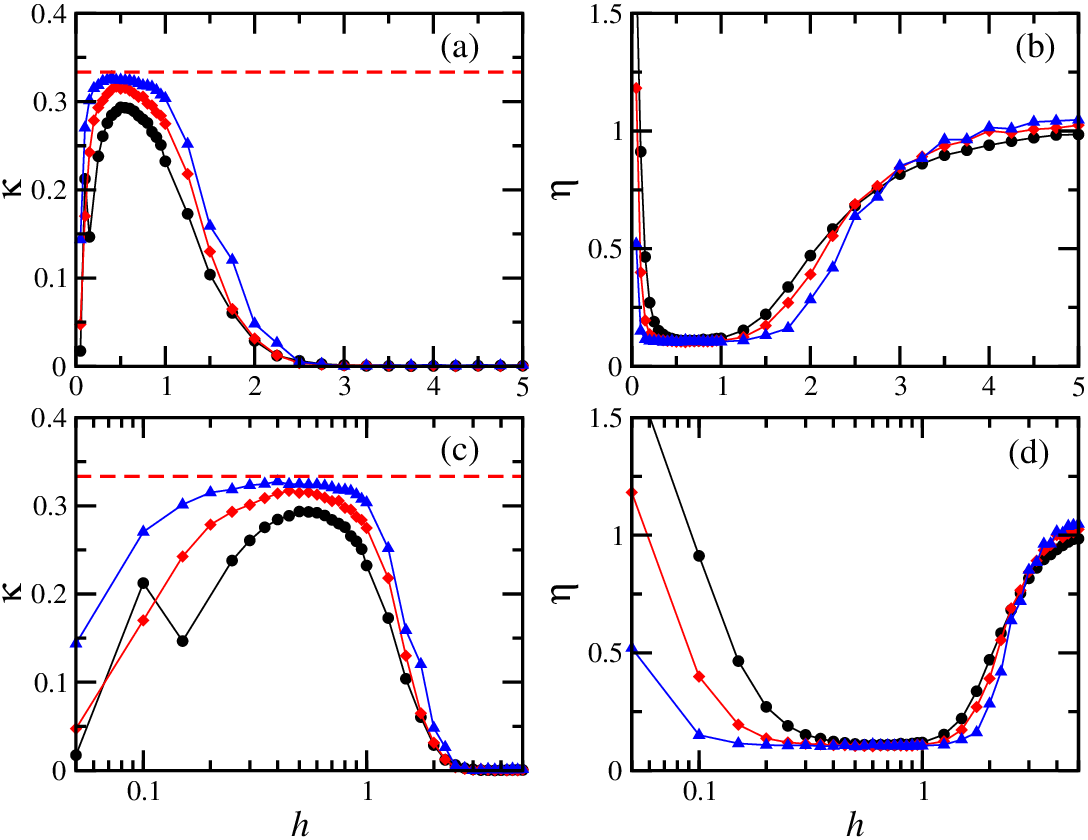}
\caption{The depth $\kappa$ of the correlation hole (a,c) and the chaos indicator $\eta$ (b,d) {\em vs.} disorder strength. The system sizes are $L = 12$ (circles), $L = 14$ (diamonds), and $L = 16$ (triangles); $\Delta=1$, $\varepsilon_{1,L}, d, \lambda=0$, ${\cal S}^z=0$. The dashed-lines in (a,c) indicate the maximum $\kappa_{FRM}=1/3$ reached by FRM. Average performed over $\sim 10^3$ different initial states corresponding to site-basis vectors with energy in the middle of the spectrum and over $10^2$ disorder realizations.}
\label{FigKappa}
\end{figure}

To quantify the depth of the correlation hole, we calculate
\begin{equation}
\kappa = \frac{\overline{P_{ini}} - \langle P_{ini}^{min}\rangle}{\overline{P_{ini}} } ,
\end{equation}
where $ \langle P_{ini}^{min}\rangle$ is the minimum value of $\langle P_{ini}(t) \rangle$. Recall from the discussions in Sec.2 (iii) that for FRMs of GOEs, $\overline{P_{ini}} \simeq 3/{\cal D}$ and $\langle P_{ini}(t)\rangle \simeq 2/{\cal D}$, so $\kappa_{FRM} =1/3$. Figures~\ref{FigKappa}  (a) and (c) show $\kappa$ as a function of the disorder strength, Fig.~\ref{FigKappa}  (a) is linear in $h$ and Fig.~\ref{FigKappa}  (b) is logarithmic in $h$. The measure $\kappa$ reaches the largest values in the chaotic region ($h\sim 0.5$), where it approaches $\kappa_{FRM}$. It decreases for $h<0.5$, as the system approaches the integrable point of the clean XXZ model, and for $h>0.5$, as the system approaches spatial localization. The depth of the correlation is therefore a general indicator of the integrable-chaos transition, which therefore captures also the spatial delocalized-localized transition.

%\vspace*{-7pt}
%\vspace*{-5pt}
 
Figure ~\ref{FigKappa} (b) and (d) shows results for the chaos indicator $\eta$ computed with the eigenvalues and defined as
\begin{equation}
\eta = \frac{\int_0^{s_0} [P(s) - P_{WD}(s)] ds}{\int_0^{s_0} [P_P(s) - P_{WD}(s)] ds}, 
\end{equation}
where $P(s)$ is the level spacing distribution, $P_P(s)$ denotes the Poisson distribution,  $P_{WD}(s)$ is the Wigner-Dyson distribution, and $s_0$ is the first point of intersection between $P_P(s)$ and $P_{WD}(s)$. When the distribution is  Poisson, $\eta \rightarrow 1$, and for the Wigner-Dyson, $\eta \rightarrow 0$. The comparison between the left and right panels makes evident the correspondence between level statistics and the long-time behavior of the survival probability. We notice, however, that while $P(s)$ detects the short-range correlations between the eigenvalues, the depth of the correlation hole detects also the long-range correlations~\cite{Ma1995}, so $\kappa$ and $\eta$ are actually complementary.

The logarithmic scale of the $x$-axis in Figs.~\ref{FigKappa} (c) and (d) draws attention to the first transition from delocalized integrable to chaos. One sees that the range of disorder strengths for which $\kappa$ is large and $\eta $ is small increases as the system size increases. This indicates that in the thermodynamic limit, this transition region may disappear and an infinitesimally small $h$ may suffice to take the system into the chaotic regime~\cite{Santos2010PRE,Torres2014PRE}. Figures~\ref{FigKappa} (a) and (b) makes more visible the second transition from chaos to spatial localization. Here also, one sees that the chaotic region gets extended, holding for larger values of $h$ as $L$ increases. This region may disappear in the thermodynamic limit, or it may reduce to a single critical point as in Anderson localization in higher dimensions, or yet it may remain a finite region. What will in fact happen to the two transition regions, integrable-chaos and chaos- localization, for $L \rightarrow \infty$ as well as the differences and similarities between the two are open questions.

\section{Conclusions}

We analyzed how the presence of level repulsion manifests itself in the dynamics of many-body quantum systems. We showed that it appears in the form of a correlation hole in the long-time evolution of the survival probability and as a minor bulge above the saturation values of the Shannon and entanglement entropies. The correlation hole detects integrable-chaos transitiosn in clean and disordered models, and by extension it is also an indicator of the spatial delocalized-localized transition in interacting systems with random onsite disorder. The correlation hole is not exclusive to the survival probability and is seen also in experimental observables, such as the density imbalance\cite{TorresARXIV2017}.

Since the correlation hole reveals the statistical properties of the spectrum from the time domain, it is advantageous to experiments that have low energy resolution and to experiments that do not have direct access to the energy levels. The hole was observed in molecules. It requires long-time coherences to be seen in experiments with cold atoms and trapped ions, although these time scales are still shorter than those needed to reach equilibration. Another pre-requisite for the observation of the correlation hole in relatively small systems is the performance of averages over initial states and disorder realizations that can smoothen the curves and thus reveal the hole.

It is worth emphasizing once again the similarities between the dynamical behavior of the Shannon entropy and the entanglement entropy in clean and disordered systems. The entanglement entropy is not only computationally more involved, but also experimentally challenging. The Shannon entropy should be a more accessible quantity to current experiments that investigate nonequilibrium quantum dynamics.

%%%%%%%%%%%%%%%%%%%% ACKNOWLEDGMENTS %%%%%%%%%%%%%%%%%%%%%
\vskip6pt

\enlargethispage{20pt}

\begin{acknowledgments}
EJTH thanks the LNS-BUAP for allowing use of their supercomputing facility and the Aspen Center for Physics hospitality, where part of this work was done. LFS thanks Antonio Garc\'ia-Garc\'ia for useful discussions. EJTH acknowledges funding from PRODEP-SEP and Proyectos VIEP-BUAP, Mexico. LFS was supported by the NSF grant No. DMR-1603418.
\end{acknowledgments}

\end{document}